\documentclass
[twocolumn,aps,prd,amsmath,showpacs,amssymb,floatfix,nofootinbib]
{revtex4}

\usepackage[T1]{fontenc}
\usepackage{CJK}                     % chinese fonts
\usepackage[dvips]{graphicx}
\usepackage{bm}                      % bold math
\usepackage{mathptmx}                % math+times fonts
\usepackage{dcolumn}                 % Align table columns on decimal point
\def\bea{\begin{eqnarray}}
\def\eea{\end{eqnarray}}
\def\be{\begin{equation}}
\def\ee{\end{equation}}
\def\non{\nonumber}

\def\eps{\varepsilon}
\def\sgm{\Sigma^-}

\def\la{\Lambda}

\def\rol{\rho_\Lambda}
\def\ros{\rho_\Sigma}
\def\ms{M_\odot}
\def\vec{\bm}

\begin{document}

\title{Hybrid stars with the Dyson-Schwinger quark model}

%\begin{CJK*}{GB}{} % Use default fonts from CJK
%\begin{CJK*}{GB}{SongMT}
\begin{CJK*}{GB}{gbsn}

\author{H. Chen (³Â»¶), M. Baldo, G. F. Burgio, and H.-J. Schulze}

\affiliation{
INFN Sezione di Catania, and Dipartimento di Fisica e Astronomia,
Universit\'a di Catania, Via Santa Sofia 64, 95123 Catania, Italy}

%\date{\today}

\begin{abstract}
We study the hadron-quark phase transition in the interior of neutron stars.
For the hadronic sector, we use a microscopic equation of state involving
nucleons and hyperons derived within the Brueckner-Hartree-Fock many-body
theory with realistic two-body and three-body forces.
For the description of quark matter, we employ
the Dyson-Schwinger approach and compare with the MIT bag model.
We calculate the structure of neutron star interiors comprising both
phases and find that with the Dyson-Schwinger model,
the hadron-quark phase transition takes place only when hyperons are excluded,
and that a two-solar-mass hybrid star is possible only if the nucleonic equation of state
is stiff enough.
\end{abstract}


\maketitle

\end{CJK*}

%===============================================================================
\section{Introduction}

The possible appearance of quark matter (QM)
in the interior of massive neutron stars
(NSs) is one of the main issues in the physics of these compact objects.
Calculations of NS structure, based on a microscopic nucleonic
equation of state (EOS), indicate that for the heaviest NS, close
to the maximum mass (about two solar masses), the central particle density
reaches values larger than $1/\rm fm^{3}$.
In this density range, the nucleon cores (dimension $\approx 0.5\;\rm fm$)
start to touch each other, and it is hard to imagine that only
nucleonic degrees of freedom can play a role.
On the contrary, it can be expected that even before reaching
these density values, the nucleons start to lose their identity,
and quark degrees of freedom are excited at a macroscopic level.

Unfortunately, it is not straightforward to predict the relevance
of quark degrees of freedom in the interior of NSs for the different
physical observables, like cooling evolution, glitch characteristics,
neutrino emissivity, and so on.
In fact, the other NS components can mask the effects coming directly
from QM.
In some cases the properties of quark and nucleonic matter are not very
different, and a clear observational signal of the presence
of the deconfined phase inside a NS is indeed hard to find.

The value of the maximum mass of a NS is one
quantity that is sensitive to the presence of QM.
If the QM EOS is sufficiently soft, the quark component is expected
to appear in NSs and to affect appreciably the maximum mass value.
In fact the recent claim of discovery of a two-solar-mass NS \cite{heavy}
has stimulated the interest in this issue.
Purely nucleonic EOS are able to accommodate masses comparable
with this large value \cite{gle,bbb,akma,zhou,zhli}.
However, the appearance of hyperons in beta-stable matter could
strongly reduce the maximum mass that can be reached by a
baryonic EOS \cite{zhli,mmy,carroll,djapo,nsesc}.
In this case the presence of nonbaryonic, i.e, ``quark'' matter
would be a possible manner to stiffen the EOS and reach large NS masses.
Heavy NSs thus would be hybrid quark stars.
In this paper, we will discuss this issue in detail.

Unfortunately, while the microscopic theory of the nucleonic EOS has
reached a high degree of sophistication, the QM EOS is poorly
known at zero temperature and at the high baryonic density appropriate for NSs.
One has, therefore, to rely on models of QM, which contain
a high degree of arbitrariness.
At present, the best one can do is to compare the predictions of
different quark models and to estimate the uncertainty of the results
for the NS matter as well as for the NS structure and mass.
Continuing a set of previous investigations using different quark models
\cite{nsquark,njl,cdm,maru,roma},
we employ in this paper the Dyson-Schwinger model (DSM) for QM
%\cite{ds},
\cite{Roberts:1994dr,Alkofer:2000wg,Maris:2003vk,Roberts:2007jh}
in combination with a definite baryonic EOS,
which has been developed within the Brueckner-Hartree-Fock (BHF)
many-body approach of nuclear matter,
comprising nucleons and also hyperons.
Confrontation with previous calculations shall also be discussed.

The paper is organized as follows.
In Sec. \ref{s:bhf}, we review the determination of the baryonic
EOS in the BHF approach.
Sec. \ref{s:qm} concerns the QM EOS according to the DSM,
comparing also with the MIT bag model for reference.
In Sec. \ref{s:res}, we present the results regarding NS structure,
combining the baryonic and QM EOS for beta-stable nuclear matter.
Section \ref{s:end} contains our conclusions.

%===============================================================================
\section{EOS of hypernuclear matter within Brueckner theory}
\label{s:bhf}

The Brueckner-Bethe-Goldstone theory is based on a linked cluster
expansion of the energy per nucleon of nuclear matter
(see Ref.~\cite{book}, chapter 1 and references therein).
The fundamental quantity of interest
in this many-body approach is the Brueckner reaction matrix $G$,
which is the solution of the Bethe-Goldstone equation,
written in operatorial form as
\be
 G_{ab}[W] = V_{ab} + \sum_c \sum_{p,p'} V_{ac} \big|pp'\big\rangle
 {Q_c \over W - E_c +i\eps}
  \big\langle pp'\big| G_{cb}[W] \:,
\label{e:g} \ee where the indices $a,b,c$ indicate pairs of baryons
and the Pauli operator $Q_c$ and energy $E_c$ characterize the
propagation of intermediate baryon pairs. The pair energy in a given
channel $c=(B_1 B_2)$ is \be
% E_{(NY)} =  Y_N(k_N) + T_Y(k_Y) + U_N(k_N) + U_Y(k_Y) \:.
 E_{(B_1B_2)} = T_{B_1}(k_{B_1}) + T_{B_2}(k_{B_2})
 + U_{B_1}(k_{B_1}) + U_{B_2}(k_{B_2})
\label{e:e}
\ee
with
$T_B(k) = m_B + {k^2\!/2m_B}$,
where the various single-particle potentials
are given by
\be
 U_B(k) =
 \sum_{B'=n,p,\la,\sgm} U_B^{(B')}(k)
% \sum_{N'=n,p} U_B^{(N')}(k) + \sum_{Y=\Sigma^-,\Lambda} U_B^{(Y)}(k)
\label{e:u}
\ee
and are determined self-consistently from the $G$ matrices,
\be
  U_B^{(B')}(k) =
  \!\!\! \sum_{k'<k_F^{(B')}} \!\!\!
  {\rm Re} \big\langle k k' \big| G_{(BB')(BB')}\!\big[E_{(BB')}(k,k')\big]
  \big| k k' \big\rangle  \:.
\label{e:ubb}
\ee
The coupled Eq. (\ref{e:g}) - (\ref{e:ubb}) define the BHF scheme
with the continuous choice of the single-particle energies.
In contrast to the standard purely nucleonic calculation,
the additional coupled-channel structure due to hyperons
renders the calculations quite time consuming.

Once the different single-particle potentials are known, the total
nonrelativistic baryonic energy density, $\eps$, can be evaluated:
\bea
 \eps \!&=& \!\!\!\!
 \sum_{B=n,p,\la,\sgm}
 \sum_{k<k_F^{(B)}}
 \left[ T_B(k) + {1\over2} U_B(k) \right] \:.
\label{e:eps}
\eea
It has been shown that
the nuclear EOS can be calculated with good accuracy in the Brueckner
two hole-line approximation with the continuous choice for the
single-particle potential, and that the results in this scheme are
quite close to the calculations which include also the three hole-line
contribution \cite{song}.

The basic input quantities in the Bethe-Goldstone equation
are the nucleon-nucleon (NN), nucleon-hyperon (NY),
and hyperon-hyperon (YY) two-body potentials $V$.
The inclusion of nuclear three-body forces (TBFs) is crucial in order to
reproduce the correct saturation point of symmetric nuclear matter.
The present theoretical status of microscopically derived TBFs is quite
rudimentary, and in most approaches semiphenomenological TBFs are used that
involve several free parameters usually fitted to the relevant data.
An important constraint is the consistency with a given two-body force, i.e.,
both two-body and three-body forces should be based on the same theoretical
footing and use the same microscopical parameters in their construction.
Recent results \cite{zhli,zuotbf} have been published within this framework,
using meson-exchange TBFs that employ the same meson-exchange parameters
as the underlying NN potential.

In this paper, we use results obtained in this manner
based on the Argonne $V_{18}$ (V18) \cite{v18},
the Bonn B (BOB) \cite{bob},
and the Nijmegen 93 (N93) \cite{n93} potentials,
and compare also with the widely
used phenomenological Urbana-type (UIX) TBFs \cite{uix}
(in combination with the $V_{18}$ potential).
We remind the reader that in our approach the TBF is reduced to
a density-dependent two-body force by averaging over the position of the
third particle,
assuming that the probability of having two particles at a given distance
is given by the two-body correlation function determined self-consistently.

In the past years, the BHF approach has been extended
with the inclusion of hyperons \cite{sch98,vi00,mmy},
which may appear at sufficiently large baryon density
in the inner part of a NS,
and lower the ground state energy of the dense nuclear matter phase.
In our work, we use the Nijmegen soft-core NSC89 NY potential \cite{nsc89}
that is well adapted to the available experimental NY scattering data
and also compatible with $\Lambda$ hypernuclear levels \cite{yamamoto,vprs01}.
Unfortunately, up to date no YY scattering data
and therefore no reliable YY potentials are available.
We therefore neglect these interactions in our calculations,
which is supposedly justified, as long as the hyperonic partial
densities remain limited.

We have previously found rather low hyperon onset densities
of about 2 to 3 times normal nuclear matter density
for the appearance of the $\sgm$ and $\la$ hyperons \cite{sch98,nsesc,vi00,mmy}
(other hyperons do not appear in the matter).
Moreover, an almost equal percentage of nucleons and hyperons are
present in the stellar core at high densities.
The inclusion of hyperons produces an EOS which turns out to be
much softer than the purely nucleonic case,
with dramatic consequences for the structure of the NS (see below).
We do not expect substantial changes when introducing refinements
of the theoretical framework,
such as hyperon-hyperon potentials \cite{vi00},
relativistic corrections, etc..
Three-body forces involving hyperons could produce a substantial stiffening
of the baryonic EOS.
Unfortunately they are essentially unknown,
but can be expected to be weaker than in the nonstrange sector.
Another possibility that is able to produce larger maximum masses,
is the appearance of a transition to QM inside the star.
This will be discussed in the next sections.

%===============================================================================
\section{Quark Phase}
\label{s:qm}

The properties of cold nuclear matter at large densities, i.e.,
its EOS and the location of the phase transition to deconfined QM,
remain poorly known.
The difficulty in performing first-principle calculations in such systems
can be traced back to the complicated nonlinear and nonperturbative nature of
quantum chromodynamics (QCD).
Therefore one can presently only resort to more or less phenomenological models
for describing QM,
and in this paper we illustrate results obtained by adopting the DSM.
A brief comparison with results from the MIT bag model will also be made.

\subsection{Dyson-Schwinger equations approach}

For the deconfined quark phase, we adopt a model based on the
Dyson-Schwinger equations of QCD,
which provides a continuum approach to QCD that can simultaneously address
both confinement and dynamical chiral symmetry breaking
\cite{Roberts:1994dr,Alkofer:2000wg}.
It has been applied with success to hadron physics in vacuum
\cite{Maris:2003vk,Roberts:2007jh,Eichmann:2008ef,Chang:2009zb},
and to QCD at nonzero chemical potential and temperature
\cite{Roberts:2000aa,Maas:2005hs,Fischer:2009wc,Chen:2008zr,Qin:2010nq}.
Recently efforts have been made to calculate the EOS for cold quark
matter and compact stars \cite{Zong:2008zzb,Klahn:2009mb}.

Our starting point is QCD's gap equation for the quark propagator
$S(p;\mu)$
at finite quark chemical potential $\mu$, which reads
\footnote{
In our Euclidean metric:
$\{\gamma_\rho,\gamma_\sigma\} = 2\delta_{\rho\sigma}$;
$\gamma_\rho^\dagger = \gamma_\rho$;
$\gamma_5 = \gamma_4\gamma_1\gamma_2\gamma_3$;
$ab = \sum_{i=1}^4 a_i b_i$;
$\bm{a}\bm{b} = \sum_{i=1}^3 a_i b_i$;
and $P_\rho$ timelike $\Rightarrow$ $P^2<0$.}
\be
 S(p;\mu)^{-1} =
 Z_2 \left[ i{\bm \gamma}{\bm p} + i \gamma_4 (p_4+i\mu) + m_q \right]
 + \Sigma(p;\mu) \:,
\label{gendse}
\ee
with the renormalized self-energy expressed as
\bea
&&\hspace{-7mm}
 \Sigma(p;\mu) =
\\&&\hspace{-4mm}
 Z_1 \int^\Lambda\!\!\!\! \frac{d^4q}{(2\pi)^4} \,
 g^2(\mu) D_{\rho\sigma}(p-q;\mu)
 \frac{\lambda^a}{2}\gamma_\rho S(q;\mu) \Gamma^a_\sigma(q,p;\mu) \:,
\non
\label{gensigma}
\eea
where $\int^\Lambda$ represents a
translationally invariant regularization of the integral,
with $\Lambda$ the regularization mass-scale.
Here, $g(\mu)$ is the coupling strength,
$D_{\rho\sigma}(k;\mu)$ is the dressed gluon propagator,
and $\Gamma^a_\sigma(q,p;\mu)$ the dressed quark-gluon vertex.
Moreover, $\lambda^a$ are the Gell-Mann matrices,
and $m_q$ is the $\Lambda$-dependent current-quark bare mass.
The quark-gluon vertex and quark wave function renormalization constants,
$Z_{1,2}(\zeta^2,\Lambda^2)$, depend on the renormalization point $\zeta$,
the regularization mass-scale $\Lambda$, and the gauge parameter.

At finite chemical potential, the quark propagator can assume a general
form with rotational covariance
\bea
 S(p;\mu)^{-1} &=&
 i {\bm \gamma}{\bm p} \;A(p^2,pu,u^2)
 + B(p^2,pu,u^2)
\non\\ &&
 +\; i \gamma_4(p_4+i\mu) \;C(p^2,pu,u^2) \:,
\label{sinvp}
\eea
where we have written $u=(\vec{0},i\mu)$.
Please note that we ignore quark Cooper pairing herein.
Diquark condensate and color superconductivity have been considered in the
DSM \cite{Yuan:2006yf,Nickel:2006vf,Nickel:2006kc},
and we will extend our analysis to that case in the future.

The kernel, Eq.~(\ref{gensigma}), depends on the gluon propagator and the
quark-gluon vertex at finite chemical potential.
However, little is known about them except at very high chemical potential,
where perturbation theory is applied.
We have to extend  to finite $\mu$ the DSM that has been
successfully applied to hadron physics at $\mu=0$.

The \emph{Ans\"atze} at zero chemical potential are typically
implemented by writing
\bea
&&\hspace{-9mm}
 Z_1 g^2 D_{\rho \sigma}(p-q) \Gamma_\sigma^a(q,p)
\non\\&&\hskip15mm
 = {\cal G}((p-q)^2) \, D_{\rho\sigma}^{\rm free}(p-q)
 \frac{\lambda^a}{2}\Gamma_\sigma(q,p) \:,
\label{KernelAnsatz} \eea wherein $D_{\rho \sigma}^{\rm
free}(k\equiv p-q) = (\delta_{\rho \sigma}-\frac{k_\rho
k_\sigma}{k^2})\frac{1}{k^2}$ is the Landau-gauge free gluon
propagator, ${\cal G}(k^2)$ is a model effective interaction, and
$\Gamma_\sigma(q,p)$ is a vertex \textit{Ansatz}.

Herein, we consider the widely used "rainbow approximation"
\be
 \Gamma_\sigma(q,p) = \gamma_\sigma \:,
\label{rainbowV}
\ee
and a Gaussian-type effective interaction \cite{Alkofer:2002bp}
\be
 \frac{{\cal G}(k^2)}{k^2} = \frac{4\pi^2D}{\omega^6} \,
 k^2\, {\rm e}^{-k^2/\omega^2} \:,
\label{IRGs}
\ee
involving two parameters $D$ and $\omega$.
This is a finite-width representation of the Munczek-Nemirovsky
model \cite{mn83} used in Ref.~\cite{Klahn:2009mb},
which expresses the long-range behavior of the
renormalization-group-improved effective interaction in
Refs.~\cite{Roberts:2007jh,Maris:1997tm,Maris:1999nt}.
Equation~(\ref{IRGs}) delivers an ultraviolet-finite model gap equation.
Hence, the regularization mass-scale $\Lambda$ can be removed to
infinity and the renormalization constants $Z_{1,2}$ set equal to one.
In this model, there is no interaction between different flavors of quarks.
Therefore, the gap equations for the different flavors are independent of
each other.
Here we consider the light flavors $u$, $d$, and $s$, neglecting heavier ones.

Usually, there exist two solutions of Eq.~(\ref{gendse}) in the
chiral limit, i.e., when $m_q=0$. One solution with nonzero quark
mass function $M(p) \equiv B(p)/A(p)$ is called Nambu
solution, and represents a phase with dynamical chiral symmetry
breaking and confinement. The other solution with zero mass function
at the chiral limit is called Wigner solution, which represents a
phase with chiral symmetry and deconfinement. The Nambu phase is
realized in vacuum, and provides the basement for describing physics
in vacuum. The phase transition to deconfinement at finite chemical
potential, without considering hadron degrees of freedom, is
investigated in Ref.~\cite{Chen:2008zr}. For strange quarks, only
the Nambu solution exists in vacuum \cite{Chang:2006bm}. The
appearance of the Wigner phase for strange quarks at finite chemical
potential will be investigated in the following.

In Ref.~\cite{Chen:2008zr}, the $\mu_q\neq 0$ \emph{Ansatz} is
specified as the same as in vacuum.
It is reasonable in the Nambu phase, but not in the Wigner phase.
In this article, we introduce a further parameter in order to study a
possible density dependence of the effective interaction.
Considering asymptotic freedom at high chemical potential,
we extend the effective interaction at finite chemical
potential in a simple manner,
\be
  \frac{{\cal G}(k^2;\mu)}{k^2} =
  \frac{4\pi^2D}{\omega^6} {\rm e}^{-\alpha\mu^2/\omega^2} k^2\,
  {\rm e}^{-k^2/\omega^2} \:,
\label{IRGsmu}
\ee
introducing the parameter $\alpha$,
which controls the rate of approaching asymptotic freedom.
In the following we will study the dependence of our results on this
parameter, marked as DS$\alpha$.
Compared to this model, the MIT bag model with noninteracting
quarks corresponds to $\alpha=\infty$.
For simplicity, we assume the
same effective interaction Eq.~(\ref{IRGsmu}) for each flavor.

The parameters $D$ and $\omega$ of the model, Eq.~(\ref{IRGsmu}),
and the quark masses can be determined by fitting meson properties
in vacuum \cite{Alkofer:2002bp}. We choose the set of parameters
$\omega=0.5\;{\rm GeV}$ and $D=1\;{\rm GeV}^2$. For simplicity, we
use current-quark masses $m_{u,d}=0$, while $m_s=115\;{\rm MeV}$ is
obtained by fitting the $K$ meson mass in vacuum, which is a little
different from the usually used value $m_s\approx150\;{\rm MeV}$ in
the MIT bag model.

The EOS of cold QM is given following
Refs.~\cite{Chen:2008zr,Klahn:2009mb}. We express the quark number
density as
\bea
 n_q(\mu) &=& 6 \int\frac{d^3 p}{(2\pi)^3} \, f_q(|\bm{p}|;\mu) \:,
\label{nqmu}
\\
 f_q(|\bm{p}|;\mu) &=&
 \frac{1}{4\pi} \int_{-\infty}^\infty \! dp_4 \,
 {\rm tr}_{\rm D}\left[-\gamma_4 S_q(p;\mu)\right] \:,
\label{nqmuf1}
\eea
where the trace is over spinor indices only.
The quark thermodynamic pressure at zero temperature can be obtained as
\be
 P_q(\mu_q) =
 P_q(\mu_{q,0}) + \int_{\mu_{q,0}}^{\mu_q}\! {\rm d}\mu \,n_q(\mu) \:.
\label{eq:pressure}
\ee

\begin{figure}[t]%..............................................................
\includegraphics[scale=0.33, bb=0 120 770 1100,clip]{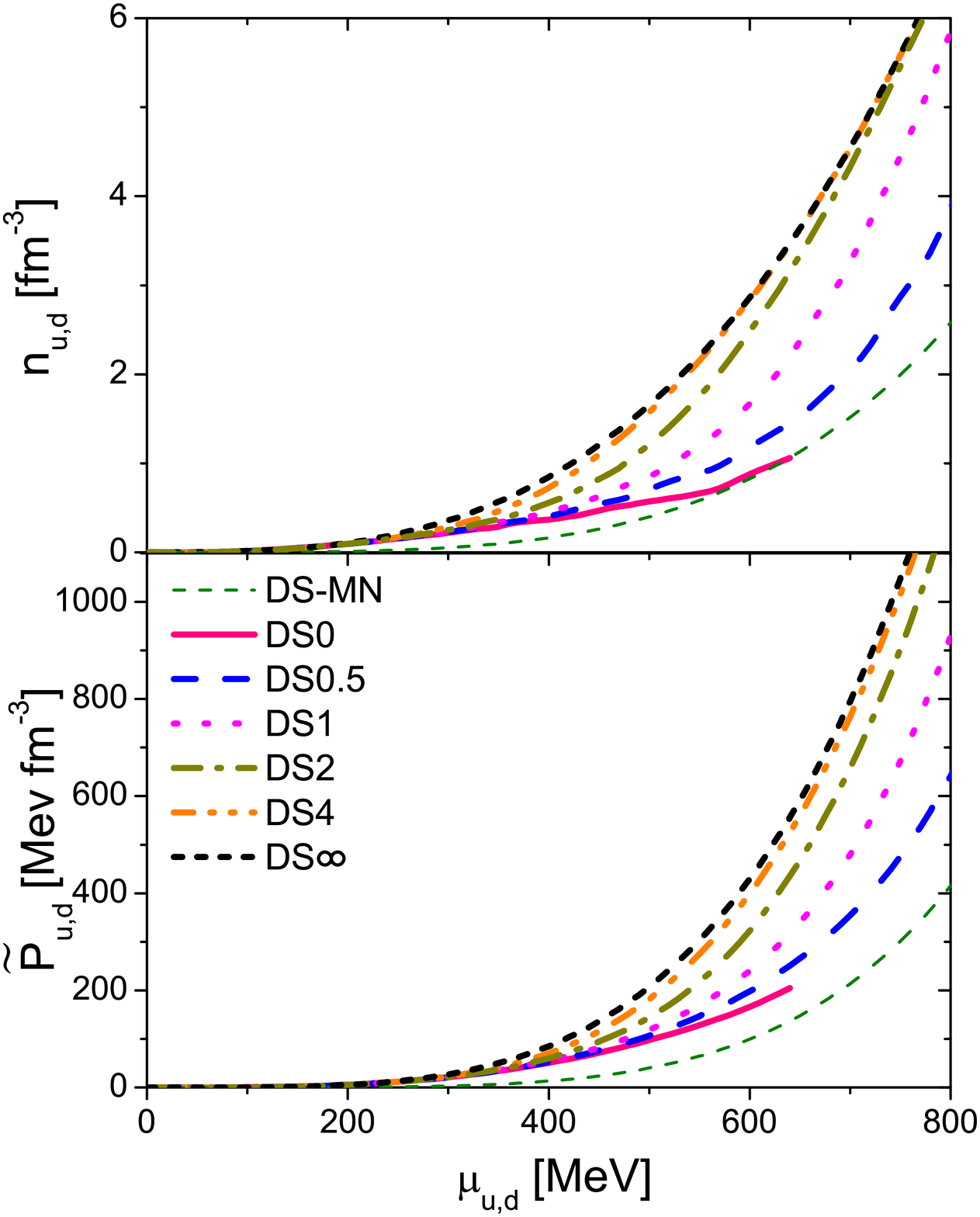}
\caption{
Quark number density (upper panel) and
pressure $\tilde{P}_q$, Eq.~(\ref{e:pqmu}), (lower panel)
for massless quarks at finite chemical potentials.
Different curves correspond to different values of the parameter
$\alpha=0, 0.5, 1, 2, 4, \infty$.
The curve denoted DS-MN
corresponds to the results in Ref.~\cite{Klahn:2009mb}.
\label{fig:light}}
\end{figure}

\begin{figure}[t]%..............................................................
\includegraphics[scale=0.33, bb=0 120 770 1100,clip]{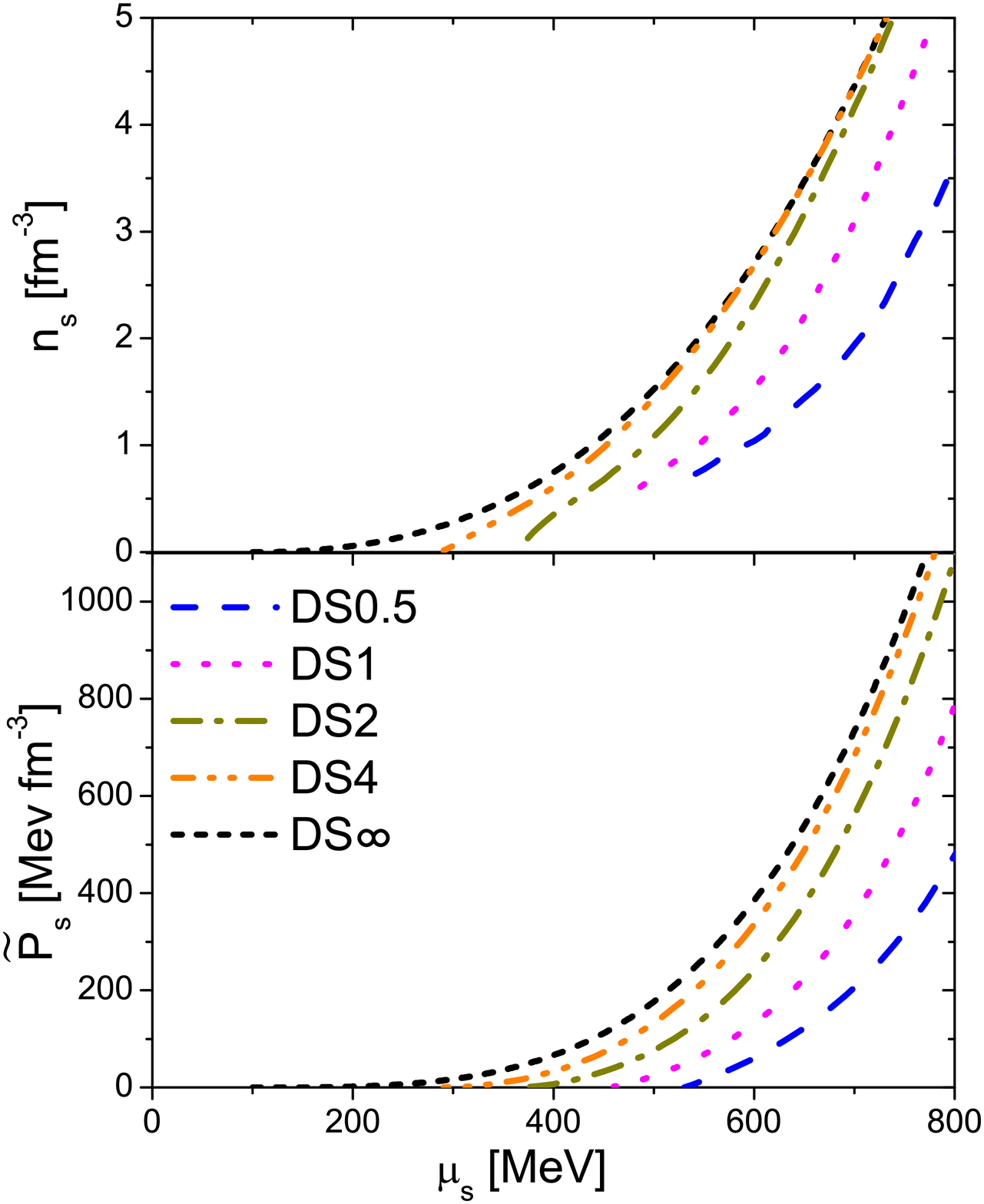} %[width=90mm,bb=12 10 290 233,clip]
\caption{
Same as Fig.~\ref{fig:light}, but
for strange quarks with $m_s=115\;{\rm MeV}$.
\label{fig:s}}
\end{figure}

The total density and pressure for the quark phase are given by
summing contributions from all flavors.
For comparison with the bag model,
we write the pressure of the quark phase as
\be
 P_Q(\mu_u,\mu_d,\mu_s) =
 \sum_{q=u,d,s}\tilde{P}_q(\mu_q) - B_\text{DS} \:,
\label{eq:PQ}
\ee
where
\bea
 \tilde{P}_q(\mu_q) &\equiv&
 \int_{\mu_{q,0}}^{\mu_q}\! {\rm d}\mu \,n_q(\mu) \:,
\label{e:pqmu}\\
 B_\text{DS} &\equiv& -\sum_{q=u,d,s} P_{q}(\mu_{q,0})\:.
 \label{e:Bdsm}
\eea
Theoretically, we can choose arbitrary values of $\mu_{q,0}$,
where the Wigner phase exists.
For massless quarks, the Wigner phase exists in vacuum and
at finite chemical potentials;
the results of $n_q(\mu_q)$ are shown in the upper panel of
Fig.~\ref{fig:light}.
Therefore, we choose $\mu_{u,0}=\mu_{d,0}=0$,
and the corresponding results of $\tilde{P}_q(\mu_q)$ are shown in the
lower panel of Fig.~\ref{fig:light}.
For strange quarks, the Wigner phase with nonzero value of $n_s$
can only be obtained with chemical potential $\mu_s$
and the parameter $\alpha$ above some thresholds \cite{Chang:2006bm};
see the upper panel of Fig.~\ref{fig:s}.
Therefore, we set $\mu_{s,0}$ as the value of the starting point of the
Wigner phase with each $\alpha$.
The single-quark number density and pressure $\tilde{P}_s$
for strange quarks are shown in Fig.~\ref{fig:s}.

Now only $B_\text{DS}$ needs to be fixed in this model.
For $u$ and $d$ quarks,
we can use the "steepest-descent" approximation \cite{haymaker:1990vm},
\be
 P[S] =  {\rm TrLn}\left[S^{-1}\right] -
 \frac{1}{2}{\rm Tr}\left[\Sigma\,S\right] \:,
\label{e:pSigma}
\ee
which is consistent with the gap equation within the "rainbow" approximation.
In vacuum we obtain the pressure difference between the Nambu phase
and the Wigner phase for massless quarks \cite{Chen:2008zr}
\be
 {\cal B} \equiv P[S_N]-P[S_W] = 45\;{\rm MeV\,fm^{-3}} \:.
\label{e:calb}
\ee
Interpreting the Nambu phase as the real vacuum with $P[S_N]=0$,
we then obtain the pressure of the Wigner phase for light
quarks in vacuum $P_{u,d}(\mu_0=0)=-45\;{\rm MeV\,fm^{-3}}$
and the effective bag constant
$B_\text{DS}^{n_f=2}=90\;{\rm MeV\,fm^{-3}}$.
However, due to the introduction of the parameter $\alpha$,
the `steepest-descent' approximation is not consistent with the gap equation
and we cannot use it to calculate $P_s(\mu_{s,0})$.
Therefore, in our model we simply set
$B_\text{DS}=90\;{\rm MeV\,fm^{-3}}$ as a parameter,
neglecting the ambiguity from $P_s(\mu_{s,0})$.

%...................................................................

%...................................................................

In summary, with respect to the MIT bag model with free quarks,
the pressure of the DSM at a fixed chemical potential is lower,
which corresponds to an increasing bag constant as
chemical potential (density) increases.
In the next section, we will show a density-dependent effective bag constant
for beta-stable QM in the interior of NSs.

\subsection{MIT bag model}

We adopt the MIT bag model assuming massless $u$ and $d$ quarks,
$s$ quarks with a current mass of $m_s=150$ MeV,
and either a fixed bag constant
$B = 90\;\rm MeV\,fm^{-3}$,
or a density-dependent bag parameter,
\bea
 B(\rho) = B_\infty
 + (B_0 - B_\infty) \exp\!\Big[-\beta \Big( \frac{\rho}{\rho_0} \Big )^2\Big ]
\label{e:brho}
\eea
with $B_\infty = 50\;\rm MeV\,fm^{-3}$,
$B_0=400\;\rm  MeV\,fm^{-3}$,
and $\beta=0.17$.
This approach has been proposed in \cite{nsquark}, and it
allows the symmetric nuclear matter to be in the pure hadronic phase at
low densities, and in the quark phase at large densities,
while the transition density is taken as a parameter.
Several possible choices of the parameters have been explored
in \cite{nsquark}, and all give a NS maximum mass in a relatively narrow
interval, $1.4\;\ms \lesssim M_{\rm max} \lesssim 1.7\;\ms$.

It has also been found \cite{nsquark,alford} that within the MIT bag model
(without color superconductivity) with a density-independent bag constant $B$,
the maximum mass of a NS cannot exceed a value of about 1.6 solar masses.
Indeed, the maximum mass increases as the value of $B$ decreases,
but too small values of $B$ are incompatible with a transition density
$\rho \gtrsim (2,\ldots,3)\rho_0$ in symmetric nuclear matter,
as indicated by heavy ion collision phenomenology.
(These baryon densities are usually reached in numerical simulations \cite{hic}
of heavy ion collisions at intermediate energies
without yielding indications of "exotic" physics.)
For a more extensive discussion of the MIT bag model,
the reader is referred to \cite{nsquark}.

%===============================================================================
\section{Results and discussion}
\label{s:res}

\subsection{Beta-stable hadronic matter}

In order to study the structure of NSs, we have to calculate
the composition and the EOS of cold, neutrino-free, catalyzed matter.
We require that the NS contains charge-neutral matter
consisting of baryons ($n$, $p$, $\la$, $\sgm$) and leptons ($e^-$, $\mu^-$)
in beta-equilibrium,
and compute the EOS in the following standard way \cite{bbb,shapiro,gle}:
The Brueckner calculation yields the energy density of
baryon/lepton matter as a function of the different partial densities,
$\eps(\rho_n,\rho_p,\rol,\ros,\rho_e,\rho_\mu)$,
by adding the contribution of noninteracting leptons to
Eq.~(\ref{e:eps}).
The various chemical potentials
(of the species $i=n,p,\la,\sgm,e,\mu$)
can then be computed straightforwardly,
\be
 \mu_i = {\partial \eps \over \partial \rho_i} \:,
\ee
and the equations for beta-equilibrium,
\be
\mu_i = b_i \mu_n - q_i \mu_e \:,
\ee
($b_i$ and $q_i$ denoting baryon number and charge of species $i$)
and charge neutrality,
\be
 \sum_i \rho_i q_i = 0 \:,
\ee
allow one to determine the
equilibrium composition $\{\rho_i(\rho)\}$
at given baryon density $\rho$ %=\rho_n+\rho_p$
and finally the EOS,
\be
 P(\rho) = \rho^2 {d\over d\rho}
 {\eps(\{\rho_i(\rho)\})\over \rho}
 = \rho {d\eps \over d\rho} - \eps
 = \rho \mu_n - \eps \:.
\ee

\begin{figure}[t]%..............................................................
\includegraphics[scale=0.8]{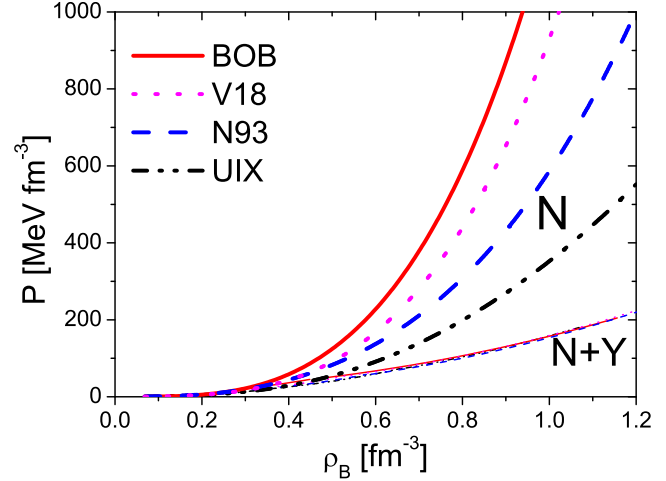}
\caption{
Pressure vs the baryon number density of hadronic NS matter.
Thick curves show results for purely nucleonic matter,
whereas thin curves include hyperons.}
\label{f:eos}
\end{figure}%...................................................................

In Fig.~\ref{f:eos}, we compare the EOS obtained in the BHF framework when only
nucleons and leptons are present (thick lines),
and the corresponding ones with hyperons included (thin lines).
Calculations have been performed with different choices of the NN potentials,
i.e., the Bonn B, the Argonne V18, and the Nijmegen N93,
all supplemented by a compatible microscopic TBF \cite{zhli}.
For completeness, we also show results obtained with the Argonne V18 potential
together with the phenomenological Urbana IX as TBFs.

We notice a strong dependence on both the chosen NN potential,
and on the adopted TBF, the microscopic ones being more
repulsive than the phenomenological force.
The presence of hyperons decreases strongly the pressure,
and the resulting EOS turns out to be almost independent of the
adopted NN potential, due to the interplay between the stiffness of the
nucleonic EOS and the threshold density of hyperons \cite{zhli}.
The softening of the EOS has serious consequences for the structure of NSs,
leading to a maximum mass of less than 1.4 solar masses \cite{zhli,mmy,nsesc},
which is below observed pulsar masses \cite{obs}.

\subsection{Quark matter in beta-equilibrium}

\begin{figure}[t]%..............................................................
\includegraphics[scale=0.75, bb=0 50 350 520,clip]{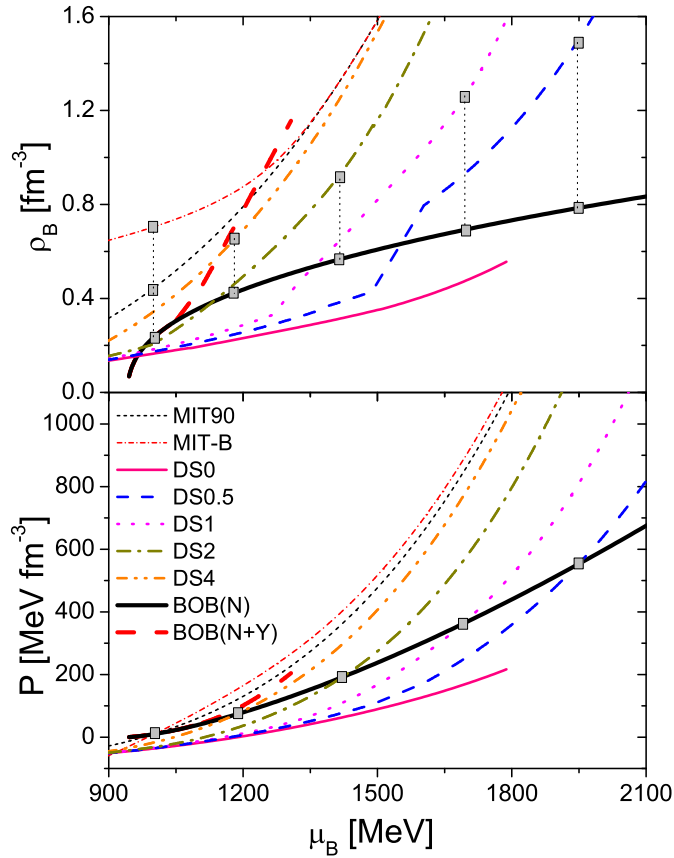}
\caption{
Baryon density (upper panel) and pressure (lower panel)
vs the baryon chemical potential of NS matter for different models.
The vertical dotted lines indicate the positions of the phase transitions
under the Maxwell construction.}
\label{f:pmu}
\end{figure}%...................................................................

In compact stars, we need to consider matter in beta-equilibrium,
\be
  d \leftrightarrow u+e+\nu_e
  \leftrightarrow u+\mu+\nu_\mu
  \leftrightarrow s \:,
\label{beta-quark}
\ee
and charge neutrality, which gives the constraints in pure QM
\bea
 && \mu_d = \mu_u+\mu_e = \mu_u+\mu_\mu = \mu_s \:,
\label{beta-quark-mu}
\\
 && \frac{2n_u-n_d-n_s}{3}-\rho_e-\rho_\mu = 0 \:.
\label{charge-quark-density}\eea

The numerical results of baryon number density $\rho_B = (n_u+n_d+n_s)/3$
and pressure versus baryon chemical potential $\mu_B = \mu_u + 2\mu_d$
are shown in Fig.~\ref{f:pmu} for the several cases discussed above.

For the hadronic case, we display results only for the Bonn-B NN potential,
which gives the stiffest EOS without hyperons,
and thus is the most favored for reaching large NS masses.
In this case the thick solid (dashed) curve indicates
results obtained without (with) hyperons.
For QM, we plot results with the MIT bag model,
with $m_{u,d}=0$, $m_s=$150 MeV, and a bag constant
$B=90\;{\rm MeV\,fm^{-3}}$,
or a density-dependent $B(\rho)$, Eq.~(\ref{e:brho}) (thin curves).
The remaining curves are results obtained with the DSM and
several choices of the model parameter $\alpha$.

The crossing points of the baryon and quark pressure curves
(marked with a square) represent the transition between baryon and QM phases
under the Maxwell construction.
The projections of these points (dotted lines) on the baryon and quark
density curves in the upper panel indicate the corresponding transition
densities from low-density baryonic matter to high-density QM.
Some qualitative considerations can be done.
In particular, we notice that the phase transition from hadronic to QM
occurs at low values of the baryon chemical potential when the MIT bag model
is used to describe the quark phase,
whereas much higher values are required with the DSM.
In some extreme cases, such as DS0, no phase transition at all is possible.
In fact, the DSM EOS is generally stiffer than the hadronic one,
and the value of the transition density is high.
We also notice that with the DSM no phase transition exists
if the hadronic phase contains hyperons.

After these indications,
we study in the following the phase transition with the more
sophisticated Gibbs construction.

\subsection{Phase transition in beta-stable matter}

A realistic model of the phase transition between baryonic and
quark phase inside the star is the Gibbs construction \cite{gle,maru,glen},
which determines a range of baryon densities where both phases coexist,
yielding an EOS containing
a pure hadronic phase, a mixed phase, and a pure QM region.
The crucial point of the Gibbs construction, as suggested by
Glendenning \cite{glen}, is that both the hadron and the quark phase
are separately charged, while preserving the total charge neutrality.
This implies that NS matter can be treated as a two-component system,
and therefore can be parametrized by two chemical potentials.
Usually one chooses the pair ($\mu_e, \mu_n$),
i.e., electron and baryon chemical potential.
The pressure is the same in the two phases to ensure mechanical stability,
while the chemical potentials of the different species are related to each
other, satisfying chemical and beta stability. As a consequence,
the pressure turns out to be a monotonically increasing function of the density.
We note that our Gibbs treatment is the zero surface tension limit
of the calculations including finite-size effects \cite{maru,yasu}.

\begin{figure}[t]%..............................................................
\includegraphics[scale=0.33, bb= 0 125 770 1110, clip]{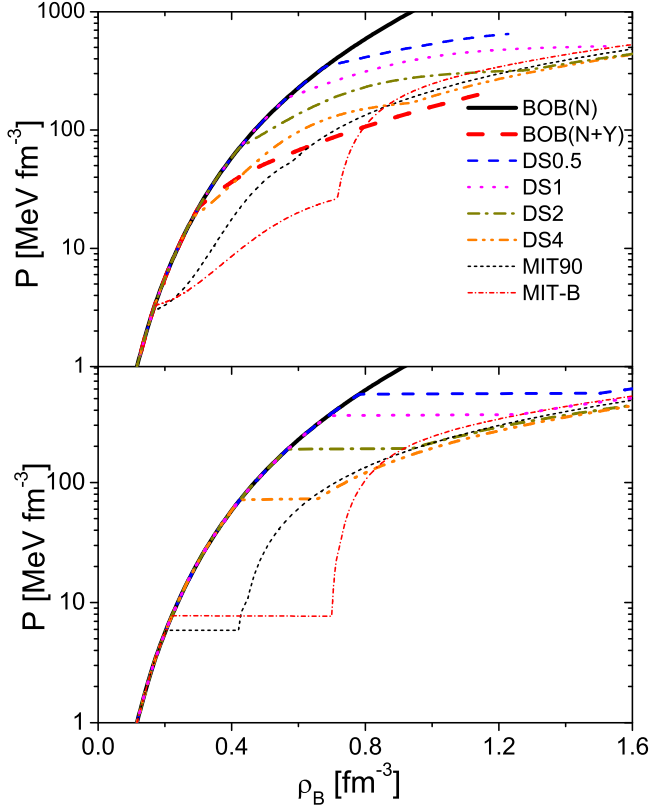}
\caption{
Pressure vs baryon density of NS matter
with the Gibbs (upper panel) or the Maxwell (lower panel)
phase transition construction for different models.}
\label{f:prho}
\end{figure} %..................................................................

The Gibbs conditions for chemical and mechanical equilibrium
between both phases read
\bea
\mu_u + \mu_e  & = & \mu_d = \mu_s \:,
\\
\mu_p + \mu_e  & = & \mu_n = \mu_\Lambda = \mu_u + 2\mu_d \:,
\\
\mu_{\Sigma^-} + \mu_p & = & 2\mu_n \:,
\\
p_H(\mu_e,\mu_n) & = &  p_Q(\mu_e,\mu_n) = p_M(\mu_n) \:.
\label{e:mp}
\eea
From these equations one can calculate the equilibrium chemical potentials
of the mixed phase corresponding to the intersection of the two surfaces
representing the hadron and the quark phase,
which allows one to calculate the charge densities $\rho_c^H$ and $\rho_c^Q$
and therefore the volume fraction $\chi$ occupied
by QM in the mixed phase, i.e.,
\begin{equation}
 \chi \rho_c^Q + (1 - \chi) \rho_c^H = 0 \:.
\label{e:chi}
\end{equation}
From this, the energy density $\eps_M$ and the baryon density $\rho_M$
of the mixed phase can be determined as
\begin{eqnarray}
 \eps_M &=& \chi \eps_Q + (1 - \chi)\eps_H \:,
\\
 \rho_M &=& \chi \rho_Q + (1 - \chi)\rho_H \:.
\label{e:mp1}
\end{eqnarray}

In Fig.~\ref{f:prho} (upper panel) we display results for the EOS
including the Gibbs hadron-quark phase transition,
using the same conventions as in Fig.~\ref{f:pmu}.
We notice that the phase transition constructed with the DSM turns out
to be quite different from the one obtained using the MIT bag model.
In the former case, if the coexistence region does exist,
it is shifted to higher baryonic density.
For comparison, the results with the simple Maxwell construction are
shown in the lower panel of the figure, the main difference being the presence
of a plateau typical of the first-order phase transitions with one conserved
charge, and the absence of a mixed phase.

\begin{figure}[t]%..............................................................
\includegraphics[scale=0.8]{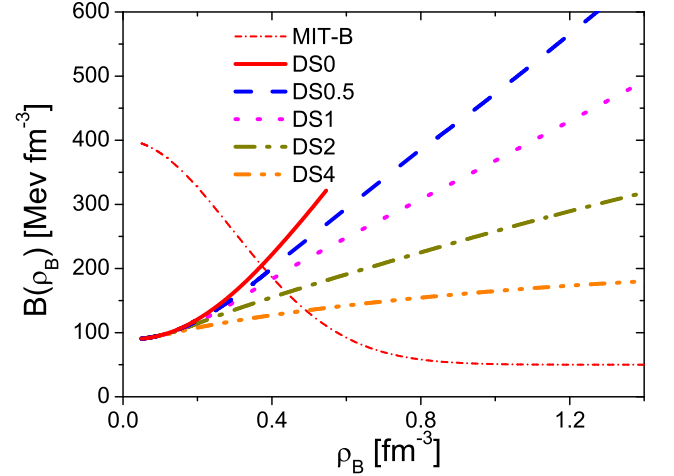}
\caption{(Color online)
Effective bag constant obtained with the MIT
model and the DSM.}
\label{f:b}
\end{figure} %..................................................................

In order to clarify the fundamental difference between MIT model on one side
and DSM on the other side,
we plot in Fig.~\ref{f:b} the effective density-dependent bag constant
\be
 B(\rho) \equiv \eps(\rho) - \eps_\text{free}(\rho)
\ee
obtained with the DSM in NS matter.
One observes that $B(\rho)$ is a monotonically increasing function
in contrast to the empirical density dependence introduced with the MIT model.
We point out that monotonically increasing bag parameters have also been
found with the color dielectric model \cite{cdm} and the Nambu-Jona-Lasinio (NJL) model \cite{bub},
whereas the density dependence in the bag model was introduced by hand
in order to delay the occurrence of the phase transition as much as possible,
see Ref.~\cite{nsquark}.
It thus appears that an increasing bag parameter is a generic feature of many
microscopic quark models.

\subsection{Neutron star structure}

We assume that a NS is a spherically symmetric distribution of
mass in hydrostatic equilibrium.
The equilibrium configurations are obtained
by solving the Tolman-Oppenheimer-Volkoff  equations \cite{shapiro} for
the pressure $P$ and the enclosed mass $m$,
\begin{eqnarray}
  {dP\over dr} &=& -{ G m \eps \over r^2 }
%\nonumber\\ && \times
  {  \left( 1 + {P / \eps} \right)
  \left( 1 + {4\pi r^3 P / m} \right)
  \over
  1 - {2G m/ r} } \:,
\\
  {dm \over dr} &=& 4 \pi r^2 \eps \:,
\end{eqnarray}
being $G$ the gravitational constant.
Starting with a central mass density $\eps(r=0) \equiv \eps_c$,
we integrate out until the density on the surface equals the one of iron.
This gives the stellar radius $R$ and the gravitational mass is then
\be
 M_G \equiv m(R) = 4\pi \int_0^R dr\; r^2 \eps(r) \:.
\ee
We have used as input the EOS obtained with the Gibbs or Maxwell construction
discussed above and shown in Fig.~\ref{f:prho}.
For the description of the NS crust, we have joined the hadronic EOS with the
ones by Negele and Vautherin \cite{nv} in the medium-density regime,
and the ones by Feynman-Metropolis-Teller \cite{fey} and
Baym-Pethick-Sutherland \cite{bps} for the outer crust.

\begin{figure}[t]%..............................................................
\includegraphics[scale=0.35]{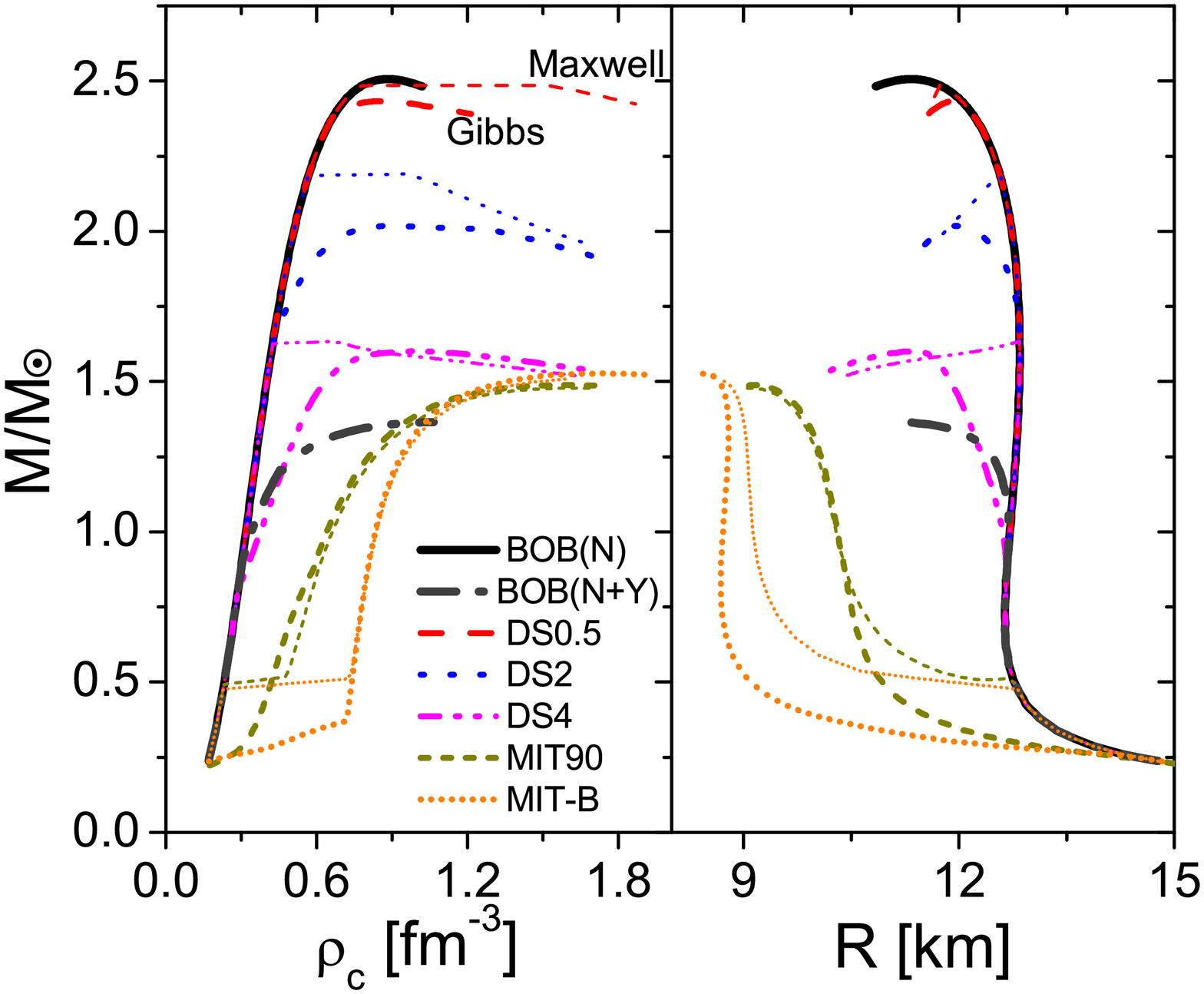}
\caption{
Gravitational NS mass vs the radius (right panel)
and the central baryon density (left panel) for different EOS
employing the BOB hadronic model.
The thick (thin) lines represent
the configurations calculated with the Gibbs (Maxwell) construction.}
\label{f:mr}
\end{figure}%...................................................................

\begin{figure}[t]%..............................................................
\includegraphics[scale=0.35]{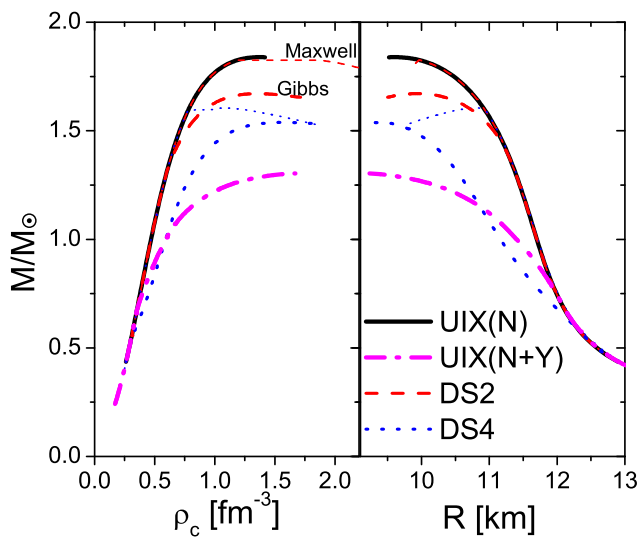}
\caption{
Same as Fig.~\ref{f:mr}, but with the UIX model.}
\label{f:mr2}
\end{figure}%...................................................................

The results are plotted in Figs.~\ref{f:mr} and \ref{f:mr2},
where we display the gravitational mass $M_G$
(in units of the solar mass $M_\odot= \rm 2 \times 10^{33}g $)
as a function of the radius $R$ and central baryon density $\rho_c$.
We present results obtained with two extreme choices of the hadronic EOS
yielding rather low or very high maximum NS masses,
namely, the UIX and BOB models, respectively.
The maximum masses are in these cases
$1.84\;\ms$ and $2.50\;\ms$ with only nucleons, and
$1.30\;\ms$ and $1.37\;\ms$ including hyperons.

The possible effects of the hadron-quark phase transition are very
different with the MIT model and the DSM: In the case of the MIT
model, the phase transition begins at very low baryon density and
thus effectively impedes the appearance of hyperons \cite{maru}.
Consequently, the resulting maximum mass of the MIT hybrid star is
$1.5\;\ms$, lower than the value of the nucleonic star, but higher
than that of the hyperon star given before.

On the contrary, with the DSM no phase transition can occur and no
hybrid star can exist if hyperons are introduced. If hyperons are
excluded, the phase transition from nucleon matter to QM takes place
at rather large baryon density. The maximum mass of the hybrid star
has a slightly smaller value than that with only nucleons,
and decreases as the density of the phase transition decreases.
For example, when the nucleonic Bonn-B NN potential is used,
the maximum mass of hybrid
stars is only a little lower than 2.5 $\ms$ with $\alpha=0.5$,
and decreases to about 2 $\ms$ with $\alpha=2$.
The same happens if
the nucleonic UIX interaction is adopted,
in which case the maximum mass of
the hybrid star cannot exceed 1.84 $\ms$.
We remind that with the value $\alpha=0$ corresponding to the
unmodified DSM without density-dependent interaction strength, no
phase transition at all is possible.
With $\alpha$ increasing to
$\infty$, we can obtain a smooth change from the pure hadronic NS to the
results with the MIT bag model.

We also notice a large difference in the structure of hybrid stars.
In fact, whereas stars built with the MIT bag model have a pure hadron phase
at low density,
followed by a mixed phase and a pure quark core at higher density,
the ones obtained with the DSM contain only a hadron phase and a mixed phase,
and probably no pure quark interior.
The scenario resembles the one obtained
within the NJL model \cite{njl,bub},
where at most a mixed phase is present, without a pure quark phase.

Another resemblance to the NJL model is the fact that with the Maxwell
construction no stable hybrid stellar configurations are obtained.
This is seen in Figs.~\ref{f:mr} and \ref{f:mr2},
where the stellar configurations
built with a Maxwell construction are represented by the thin lines,
and are characterized by a plateau.
The sudden onset of the high-density quark phase renders the star unstable,
with maximum mass values
(of purely hadronic stars)
which turn out to be
slightly larger than the ones calculated with the Gibbs construction.

A clear difference between the two models exists as far as the radius
is concerned. Hybrid stars built with the DSM are characterized by
a larger radius and a smaller central density,
whereas hybrid stars constructed with the MIT bag model are more compact.

%===============================================================================
\section{Conclusions}
\label{s:end}

We have investigated the capability of the DSM for QM to provide
hybrid NS configurations in combination with a microscopic hadronic
EOS obtained within the BHF formalism including also the appearance
of hyperons.
We found that the unmodified DSM does not allow a
hadron-quark phase transition, thus requiring the introduction of an
empirical density dependence of the quark interaction strength.
Even in this case, however, the early appearance of hyperons in hadronic
matter inhibits the phase transition.
Only in the fictitious case of restricting to pure nucleon matter,
a phase transition at large baryon density is possible,
and a hybrid star with 2 $\ms$ is only allowed if the nucleonic EOS
is stiff enough to produce a NS with 2 $\ms$.
Furthermore, stable hybrid stars are only obtained with the Gibbs
phase transition construction, but not with the Maxwell construction.

These features of the DSM are very different from MIT-type quark models,
and we have attributed it to the interaction between quarks in the DSM,
which can be represented as a density-dependent bag constant.
We have found a different density dependence of the effective bag constant
in the DSM and the MIT quark model.

\section*{Acknowledgments}

This work was partially supported by CompStar,
a Research Networking Programme of the European Science Foundation,
and by the MIUR-PRIN Project No. 2008KRBZTR.

%%%%%%%%%%%%%%%%%%%%%%%%%%%%%%%%%%%%%%%%%%%%%%%%%%%%%%%%%%%%%%%%%%%%%%%%%%%%%%%%

\end{document}